# Observational Evidence for Weak Gravitational Lensing by Large-Scale Structure

Jens Verner Villumsen

*Max Planck Institut für Astrophysik,*
*Karl Schwarzschild-Str. 1, Postfach 15 23, 85740 Garching, Germany*
*jens@mpa-garching.mpg.de*

4 July 1995

**ABSTRACT**
A single very deep 10' field has been imaged to $r \sim 26$. There are 2682 galaxies with $23 \leq r \leq 26$ with a size significantly larger than the seeing disk in the field. After correcting for telescope aberrations, possible guiding errors, and signal degradation due to seeing, I find a polarization signal of amplitude $|p| = 2.7\%$, with an uncertainty of 1.2% (95% confidence limit). For the 1773 galaxies with $23 \leq r \leq 25$, the measured amplitude is $2.4\% \pm 1.2\%$ (95% confidence limit).

If this observed polarization is due to gravitational lensing by large scale structure, cosmological inferences can be made. The polarization amplitude would indicate that $\Omega_0 \times \sigma_8 \approx 1$.

The amplitude probability distribution is a Rayleigh distribution with variance $\sigma_p^2$ so the probability $P(1/2 < |p|/\sigma_p < 2) = 75\%$, and $P(1/3 < |p|/\sigma_p < 3) = 93.5\%$. The measurement of the polarization amplitude in a single field can be used to constrain cosmological models since $\sigma_p \propto \Omega_0 \times \sigma_8$.

## 1 INTRODUCTION

A central question of modern cosmology is the degree to which galaxies trace the underlying mass distribution of the universe. Large-scale inhomogeneities in the mass distribution have been inferred from the dynamics of galaxies (eg. Faber & Burstein 1988, Bertschinger & Dekel 1989). However, the determination of peculiar motions of galaxies is fraught with difficulties such as sensitivity to small errors in distance indicators and inhomogeneous Malmquist bias. As a result, observations that seem to indicate large-scale coherent flows of galaxies remain controversial.

Weak gravitational lensing of distant galaxies, equivalent to a systematic induced image ellipticity without multiple imaging, is an alternative probe of the large-scale mass distribution of the universe (eg. Kristian 1967, Gunn 1967). In this case photons are used as a probe of the potential instead of galaxies. Essentially, when light rays from a distant galaxy pass by an overdense region in the mass distribution, the observed images are slightly elongated tangentially with respect to the center of the perturbation. Such an effect has been observed in a number of rich clusters of galaxies, eg. Tyson, Valdes & Wenk(1990), Kaiser & Squires (1993). Underdensities in the mass distribution produce images with radial elongation. Independent perturbations along the line of sight add stochastically (Blandford et al. 1991 [BSBV], Miralda-Escudé 1991, Kaiser 1992, Villumsen 1995).

Early searches for weak lensing outside the region of rich clusters resulted in unsurprising non-detections (Kristian 1967, Valdes, Tyson & Jarvis 1983). More recently, from a single deep CCD frame of a blank field Mould et al. (1994) [M-94] attempted to measure the mean image polarization, $p$, of 4363 galaxies with magnitudes $23 \leq r \leq 26$. Inside a circular aperture of radius 4.8', $p = 0.01 \pm 0.01$ was found by M-94.

As noted in M-94, there exists in the data a non-circular point spread function (PSF), possibly due to a telescope guiding error. Such an effect will lead to an artificially induced image ellipticity that can mimic a cosmological gravitational lensing signature. The artificial signal is, however, present in both the galaxies and the stars (which will not be gravitationally lensed) and will be of least importance for large galaxy images. It can, in principle, be removed by an extrapolation to infinite image size and an attempt to do this was carried out in M-94. In this *Letter* a re-analysis of the M-94 data is presented, focusing on the removal of the PSF from the galaxy images and optimal weighting of the data to find the mean of the distribution of image polarizations. Provided it is possible to estimate the PSF from the unsaturated stellar images, a mean image polarization of $p = 2.7\% \pm 1.2\%$ (95% confidence bounds) is found.

The theory of the removal of the PSF to obtain the true image orientation is outlined in §2. In §3 the observational data and limits on classical aberrations induced by the telescope and imager are discussed. In §4 the results of §2 are applied to the M-94 data using an optimal weighting scheme based upon the observed distribution of image orientations. A discussion of the implications of the results of §4 is given in §5.

## 2 IMAGE POLARISATION AND PSF REMOVAL

In the notation of BSBV the complex orientation of an image (galaxy or star) is defined as $\chi = (I_{22} - I_{11} - 2iI_{12})/(I_{11} + I_{22})$, where $I_{ij}$ are the image second moments. The modulus of $\chi$ is $(a^2 - b^2)/(a^2 + b^2)$ where $a^2$ and $b^2$ are the principal second moments of the intensity. To linear order the modulus of $\chi$ for an image is equal to its ellipticity as defined conventionally. The phase of $\chi$ is twice the position angle, $\phi$.

The effect I seek is the mean polarization, a net ellipticity in the galaxy images induced by weak lensing by large-scale structure. In the limit of mildly elliptical images, as will be assumed, the mean polarization is $p = \langle \chi \rangle$, where the brackets $\langle \ \rangle$ denote an average over an area of sky large enough to contain many galaxy images but smaller than the size of the large-scale structure being probed. Assuming $p \simeq 0.03$, as would be typical of an unbiased CDM universe with $\Omega_0 = 1$ (eg. M-94), $\sim 1000$ galaxy images are needed to measure this level of image polarization in the absence of systematic errors (eg. BSBV).

To evaluate $\langle \chi \rangle$, the distribution of the true image orientations must be known. In the limit of noise-free data, the observed $\chi$ for an image, $\chi^o$, is equal to the true $\chi$, $\chi^t$. For observational data the primary difficulties in determining $\chi^t$ are effects of the atmosphere (eg. seeing) and trailing of the images due to imperfect guiding. For sufficiently large images the effects of guiding errors and seeing can be ignored. However, for typical images seeing diminishes the eccentricity ("circularization") and guiding errors induce artificial eccentricity which can mimic a lensing signal. Possible problems with matching of frames (i.e. linear and rotational offsets) can be characterized by an asymmetric PSF. Such effects can be modeled as a convolution of the true images with a PSF and here we outline the procedure for removing the effect of the PSF from images comparable to or larger than the isophotal areas of the unsaturated stellar images. Using the unweighted image second moments, I show that if we know the second moments of the PSF (as characterized by the unsaturated stellar images) the mapping from $\chi^t$ to $\chi^o$ is straightforward.

Consider the effect of the PSF on a single image. The light profile of the true image is $\mu(x_1, x_2)$ and the PSF is characterized by a function $\mathcal{S}(x_1, x_2)$. No specific assumptions about the forms of $\mu$ and $\mathcal{S}$ are made other than that they both have unit integral and zero dipole moment. From the convolution of $\mu$ and $\mathcal{S}$ an expression for the true unweighted image second moments, $I_{ij}^t$, in terms of the observed image second moments, $I_{ij}^o$, and the second moments of the PSF, $\Delta_{ij}$, can be derived.

A photon that would have arrived at the detector at $(x_1, x_2)$ in the absence of the PSF has a probability density $\mathcal{S}(\Delta x_1, \Delta x_2)$ of arriving at $(x_1 + \Delta x_1, x_2 + \Delta x_2)$. So the photon that would have contributed $x_i x_j$ to the second moments instead contributes the term

$$
\begin{aligned}
x_i x_j &\rightarrow \int d^2 \Delta x \ (x_i + \Delta x_i)(x_j + \Delta x_j) \ \mathcal{S}(\Delta x_1, \Delta x_2) \\
&= \int d^2 \Delta x \ (x_i x_j + x_i \Delta x_j + x_j \Delta x_i + \Delta x_i \Delta x_j) \\
&\quad \times \mathcal{S}(\Delta x_1, \Delta x_2) \ = \ x_i x_j + \Delta_{ij}. \quad (1)
\end{aligned}
$$

Integrating over the entire image the result is

$$I_{ij}^o = \int d^2 x \ (x_i x_j + \Delta_{ij}) \ \mu(x_1, x_2) = I_{ij}^t + \Delta_{ij}. \quad (2)$$

That is, the observed image second moments are simply a sum of the true second moments (i.e. the moments one would measure in the absence of the effect of the PSF) and the second moments of the PSF. From this the observed image orientation, $\chi^o$, can be parametrized in terms of the true image orientation, $\chi^t$, and the second moments of the PSF:

$$
\begin{aligned}
\chi^o &= \frac{I_{22}^o - I_{11}^o - 2iI_{12}^o}{I_{22}^o + I_{11}^o} \\
&= \frac{I_{22}^t + \Delta_{22} - I_{11}^t - \Delta_{11} - 2iI_{12}^t - 2i\Delta_{12}}{I_{22}^o + I_{11}^o} \\
&= \chi^t + \frac{-\chi^t(\Delta_{22} + \Delta_{11}) + \Delta_{22} - \Delta_{11} - 2i\Delta_{12}}{I_{22}^o + I_{11}^o}. \quad (3)
\end{aligned}
$$

I define a variable $T = 1 - (\Delta_{11} + \Delta_{22})/(I_{11}^o + I_{22}^o)$ for an image and the complex orientation of the PSF as $\chi^S = (\Delta_{22} - \Delta_{11} - 2i\Delta_{12})/(\Delta_{22} + \Delta_{11})$. With this parametrization the observed image orientation, $\chi^o$, is a simple linear function of $T$, $\chi^t$, and $\chi^S$

$$\chi^o(T) = \chi^S + \left(\chi^t - \chi^S\right) T. \quad (4)$$

Thus for $T \approx 0$ (i.e. a very small galaxy) the observed $\chi$ for an image is just that of the PSF while for a very large galaxy, $T \approx 1$, $\chi$ is unaffected by the PSF and therefore $\chi^o = \chi^t$. Furthermore, the observed $\chi$ is a linear function of $T$. This result follows from simple parametrizations and does not require any special assumptions about the image and PSF profiles. Additionally, it is not necessary to assume the PSF results from seeing and guiding errors. Neither is it necessary to assume galaxies follow a standard profile. In principle eqn. 4 allows us to infer the true $\chi$ for a single image; however due to the presence of noise this is an unstable procedure and will not be attempted.

For given values of $T$, I average $\chi$ over the appropriate images and find the mean image polarization as a function of $T$ is

$$p(T) = \chi^S + \left(p^t - \chi^S\right) T. \quad (5)$$

Thus, if I measure the image polarization as a function of $T$ and I also know $\chi^S$, then the true mean polarization of the galaxies, $p^t$, is the linear extrapolation of $p(T)$ to $T = 1$.

In observational data the brightness is not a continuous function, but rather the data is in the form of photons per pixel. This will induce an error in the second moments, however, as long as an image is well sampled, as in the present data, this error is negligible. A further problem is that the unweighted second moments are unstable to the presence of noise. To counter that instability, the data is usually cut at some surface brightness level and only contiguous pixels with a surface brightness above the critical level are counted in the second moments. This will lead to an error in the mapping from true $\chi$ to observed $\chi$ (Eq. 4). For a given data set it is necessary, through Monte Carlo simulations, to measure how accurate this mapping is in the presence of noise. For the current data set, with a magnitude limit of $r < 26$, this mapping is accurate within the noise in the data and does not induce spurious polarizations.

## 3 OBSERVATIONAL DATA

The imaging data used are of a single $9.6' \times 9.6'$ blank field centered on $\alpha(1950) = 17^h 21^m 07^s$ $\delta(1950) = +49°52'21''$, taken in Gunn $r$. The data were acquired with the COSMIC imaging spectrograph (Dressler *et al.* 1995) on the 5-m Hale telescope during periods of good seeing ($0.7''$–$0.9''$). The final stacked image consists of a total of 24.0 ksec integration, has a 1$\sigma$ surface brightness limit of $\mu_r = 28.8$ mag arcsec$^{-2}$, seeing of $0.87''$ FWHM, and a total area of 90.1 arcmin$^2$. An object catalogue was created from this frame using FOCAS (Valdes 1982) and contains $\sim 6600$ objects brighter than the 80% completeness limit of $r = 26.2$. The reduction of the data to a catalogue of detected objects is detailed in M-94, which also estimates the mean redshift of galaxies in the range $23 \leq r \leq 26$ to be $z \approx 3/4$.

An obvious source of spurious ellipticity in the images is classical aberration due to the primary mirror and imager. From several offset images of the globular star cluster M3, M-94 placed limits on the amount of classical aberration in the data. The only significant aberration found was distortion (*i.e.* the classic "barrel" or "pincushion"), corresponding to a radial image displacement of $-2.6''$ in the frame corners and $-0.9''$ at the centers of the frame edges. Individual frames were geometrically remapped prior to stacking, otherwise an artificial polarization of $\sim (\Delta\theta/10'')^2$ percent would have been created in the combined frame, where the frame to frame displacement is $\Delta\theta \sim 10''$–$30''$. As in M-94, due to the significant remapping required in the frame corners, analysis of the data is limited to objects whose centroids are within a circular aperture of radius $4.8'$, centered on the chip. Results of direct measurements of the classical aberrations from the star fields agree well with a ray trace analysis of the COSMIC optical design using the ZEMAX code and the spurious polarization induced by all classical aberrations is estimated to be $p \lesssim 0.01$. The observed two point correlation function of the shear $C_{pp}(\theta)$, see M-94, Fig. 2, can be used to constrain the aberration induced polarization. This induced polarization would lead to an anticorrelation in $C_{pp}(\theta)$ on large scale which is not observed.

It is also possible for atmospheric effects to produce spurious ellipticities. From the width of the $r$ filter, M-94 estimated that atmospheric refraction contributes a uniform polarization of at most 0.006 in the individual frames and the net induced ellipticity in the combined frame is negligible.

Sensitivity variations across the field due to inadequate flatfielding can also induce spurious polarizations. However, the variations would have to be much larger than the estimated flatfielding uncertainty to significantly influence the results.

## 4 OBSERVED MEAN IMAGE POLARIZATION

Here the results of §2 are applied to the M-94 data and an estimate of $p^t$ is obtained. The stars in the frame provide a measure of the PSF since they are intrinsically point-like objects that are not subject to gravitational lensing. The second moments of the unsaturated stellar images are calculated in the same manner as those of the galaxies and provide measures of $\chi^S = (-0.008 \pm 0.005) + (0.017 \pm 0.005)i$ and $\Delta_{22} + \Delta_{11} = 5.9 \pm 0.1$ (pixel units). These numbers are the crucial ingredients for the PSF removal. In a least squares fit, including linear and quadratic terms, to the PSF as a function of position, no statistically significant variations of the PSF across the field were found.

Throughout the analysis all galaxy images with $23 \leq r \leq 26$, isophotal area larger than the mean isophotal area of the unsaturated stars, and whose centroids are within $4.8'$ of the center of the chip are used. There are 2682 galaxies that satisfy these criteria.

A single image can be seen as a measurement of the polarization with errors given by the intrinsic eccentricity of the source plus the noise associated with the measurement and I wish to find the optimum estimator for the polarization. To do this I define $p(T) = \text{Re}(\langle \chi(T) \rangle) + i\,\text{Im}(\langle \chi(T) \rangle) \equiv p_x(T) + i p_y(T)$. The 2682 galaxy images were divided into 10 equally-spaced bins in $T$ and $\langle \chi_x(T) \rangle$ and $\langle \chi_y(T) \rangle$, together with $\chi^S$, were used to determine $p_x(T)$ and $p_y(T)$ independently as follows.

In order to use a maximum likelihood (ML) estimator of $p_{x,y}(T)$ the distribution $f$ of $\chi_{x,y}$ for a given value of $T$ must be known. In Fig. 1 the distribution of $|\chi_y|$ for galaxy images with $0 < T < 1$ is shown. From this figure, for $T \lesssim 0.8$ the distribution is nearly exponential with a turnover near zero and is flat for larger $T$. Similar results are obtained for the distribution of $|\chi_x|$. The mean eccentricity is observed to be an increasing function of $T$. This is due to two effects, the larger images are less affected by the circularizing effects of seeing, and for a given image size the eccentricity is an increasing function of $T$.

In an ML estimation of the polarization, the optimum weightfunction $W_{ML}$ is $W_{ML}(x) \propto -[(df/dx)/(f(x) \cdot x)]$. The ML estimator then suggests weighting of the data to find $\langle \chi_{x,y} \rangle$ where the weight function is of the form $W_{ML}(\chi_{x,y}) = [1 + (\chi_{x,y}/\delta)^2]^{-1/2}$. The uncertainty, $\delta$, in individual measurements of $\chi_{x,y}$ is estimated to be $\sim 0.07$. Such a weighting would indicate that the information is in the rounder images. However, the ML estimator may not be the optimal estimator since, in order to apply it, the second derivative of the distribution function $f$ must be well-behaved. This is not true for a peaked distribution as in Fig. 1, in which case it is better to search for the mode of the distribution. For this reason I parametrize a weighting function $W_n(\chi_{x,y}) = [1 + (\chi_{x,y}/\delta)^2]^{n/2}$ and determine $p(T)$ using several choices of $n$. More negative values of $n$ correspond to a stronger weighting of the data towards round objects and $n = 0$ corresponds to no weighting. The mean and variance of the $\chi_{x,y}$ distributions are determined according to the standard definition of the weighted mean. For large values of $T$ the distributions of $\chi_{x,y}$ are no longer exponential and no weighting of the data is done to determine $\langle \chi_{x,y} \rangle$ in the largest $T$ bin.

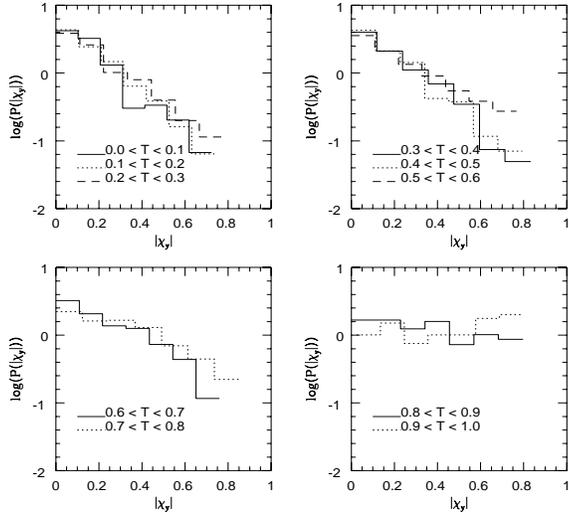

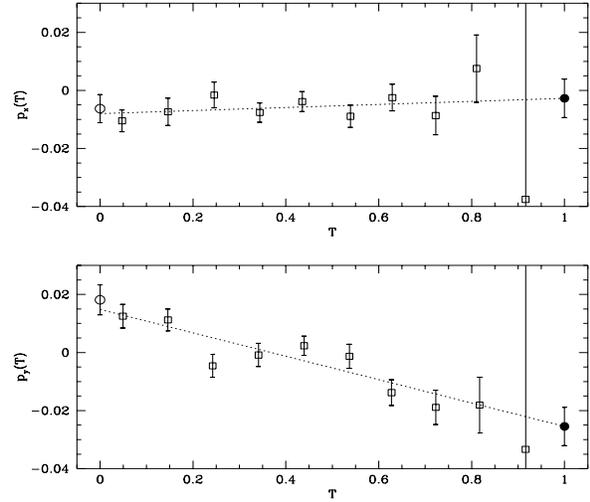

**Figure 1.** Distribution of $|\chi_y|$ in 10 bins of $T$. The ordinate is logarithmic.

**Figure 2.** Linear fits to $p_{x,y}(T)$ for $n = -2$, $\delta = 0.07$. The open squares indicate galaxy data points, the open circles are the PSF data point. The filled circles indicate the best fit values for the true polarization. All error bars are formal 1-$\sigma$ error bars.

From eqn. 5, the real and imaginary components of $p^t$, $p_{x,y}$, are estimated independently by standard weighted linear least squares fits to the observed components of $p(T)$ and $\chi^S$. An important point is that, when determining $\langle \chi_{x,y}(T) \rangle$, the appropriate value of $\chi$ in $W_n$ is $\chi^t$ not $\chi^o$. Thus, best estimates of $p_{x,y}$ are determined in an iterative manner. For a given value of $n$, I begin by assuming $\chi^t = \chi^o$, use $\chi^o$ as $\chi$ in $W_n$ and determine first estimates of $p_{x,y}$ as the extrapolations to $p(T = 1)$ of the least squares fits to the components of $p(T)$. The first estimate of $p_{x,y}$ is then used to estimate $\chi^t$ from eqn. 5, which is then used as $\chi$ in $W_n$, and new estimates of $p_{x,y}$ are obtained using least squares and extrapolating to $p(T = 1)$. From the new estimates of $p_{x,y}$, new estimates of $\chi^t$ are obtained, and the entire procedure is repeated until convergence is reached (typically less than 10 iterations). The procedure has, from Monte Carlo simulations, proven to be stable and, from the data itself, has proven to be insensitive to the starting assumptions of $p_{x,y}$. In Fig. 2, the final fits to $p_{x,y}(T)$ for the case of $n = -2$ and $\delta = 0.07$ are shown. The PSF data points are located at $T = 0$ and were included in each iteration of the linear fits to $p(T)$. The points at $T = 1$ indicate the inferred mean polarization and the error bars are the formal estimates from the least squares fit. The straight lines in Fig. 2 are clearly good fits but they are, in reality, too good since the procedure used to determine $p(T)$ introduces correlations between the data points, resulting in an underestimate of the errors in $p_{x,y}$. To obtain more realistic estimates of the errors in $p_{x,y}$, the galaxy and star images were bootstrap resampled 1000 times and the corresponding best-fit $p_{x,y}$ determined. Results for $n = (0, -1, -2, -3)$ are shown as scatter plots in

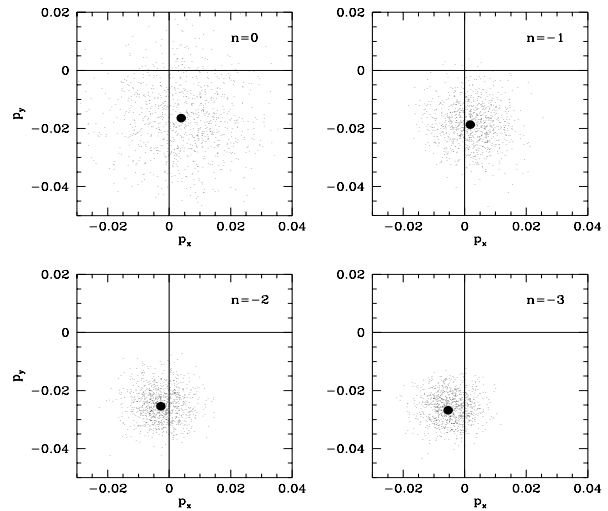

**Figure 3.** Scatter plots of best fit true polarization for 1000 bootstrap samples of the data, $23 < r < 26$, for $n = (0, -1, -2, -3)$, $\delta = 0.07$. Large dot indicates best fit value of polarization.

Fig. 3 and I take the variance in $p_{x,y}$ to be the variance as determined from the resamplings.

The optimal value of $n$ to use in the weighting of the data to determine $\langle \chi_{x,y} \rangle$ has not yet been addressed. A useful criterion for the optimal value of $n$ is that which minimizes the variance in $\chi^t$. From Fig. 3, the variance in $p_{x,y}$ is smallest for $n = -3$. In Table 1 best estimates of the true polarization and its associated error are listed for $W_n$ with $n = 0, 1/2, -1, .., -4$. For $n < -2$ the polarization and the error are nearly independent of $n$ and the formal best value of the mean polarization is $(p_x, p_y) = (-0.5\pm0.5, -2.7\pm0.6)$. This error estimate includes the error induced by the uncertainty in what is the optimal weighting scheme. When this analysis is repeated for the 1773 galaxies with $23 < r < 25$, very similar results are obtained, see Table 1. This indicates that the signal is not generated exclusively by galaxies near the completeness limit. From Table 1 it is seen that the error bars for the polarization determinations with different weighting schemes overlap.

For $r > 25$ there are very few images significantly larger than the seeing disk. This means that most of the statistical significance of the polarization measurement comes from the brighter images, *i.e.* $23 < r < 25$, and that it is not possible to derive the true polarization from the faintest sources, *i.e.* $25 < r < 26$, only. This indicates that going to very faint magnitudes is only meaningful in exceptional seeing conditions.

|  | $23 < r < 26$ | | | $23 < r < 25$ | | |
|---|---|---|---|---|---|---|
| $n$ | $p_x$ | $p_y$ | Error | $p_x$ | $p_y$ | Error |
| 0 | +0.4 | -1.6 | 1.2 | -0.2 | -1.4 | 1.2 |
| -0.5 | +0.3 | -1.7 | 0.9 | -0.2 | -1.7 | 0.9 |
| -1.0 | +0.2 | -1.9 | 0.7 | -0.2 | -2.0 | 0.8 |
| -1.5 | +0.0 | -2.2 | 0.6 | -0.3 | -2.3 | 0.7 |
| -2.0 | -0.3 | -2.5 | 0.6 | -0.4 | -2.4 | 0.6 |
| -2.5 | -0.5 | -2.7 | 0.5 | -0.4 | -2.4 | 0.6 |
| -3.0 | -0.5 | -2.7 | 0.5 | -0.5 | -2.3 | 0.6 |
| -3.5 | -0.5 | -2.6 | 0.5 | -0.5 | -2.2 | 0.5 |
| -4.0 | -0.5 | -2.5 | 0.5 | -0.5 | -2.1 | 0.5 |

Table 1: Table of measured "true" polarizations $(p_x, p_y)$ (in %) with error estimate for different values of the weighting parameter $n$. Columns 2-4 are for the magnitude range $23 < r < 26$, while columns 5-7 are for the magnitude range $23 < r < 25$.

The weighting scheme is useful for finding the mean polarization across the field. However, this same weighting scheme would lead to an artificially flat $C_{pp}(\theta)$ if applied uniformly across the field. There are not enough galaxies in the field to allow the measurement of the variations of the polarization across the field using an optimal weighting scheme. $C_{pp}(\theta)$ then has to be determined from the unweighted data and the error bars are too large to make cosmological inferences.

## 5 DISCUSSION

I have measured a statistically significant polarization in a single field. The most controversial aspect of the data reduction is the PSF removal. Under the assumptions that systematic errors have been dealt with properly and that the polarization is induced by gravity on cosmological scales I can make theoretical inferences. Since the variance of the polarization scales as the first moment of the power spectrum of density fluctuations with a low pass filter, the main contribution to a polarization signal will occur near the scale where the slope of the power spectrum is -1. For a CDM, or HDM power spectrum (Bardeen *et al.* 1986) the mean polarization for a 10' field, where the sources are at a typical redshift $z \approx 3/4$, will be generated on scales, $L_{\text{CDM}} \equiv 2\pi/k_{\text{CDM}} \approx 20\text{Mpc}/(\Omega_0 h)$, and $L_{\text{HDM}} \equiv 2\pi/k_{\text{HDM}} \approx 30\text{Mpc}/(\Omega_0 h)$. Here $h$ is the Hubble constant in units of $100\text{km s}^{-1}\text{Mpc}^{-1}$.

Villumsen (1995), §3.4, shows that the most likely polarization amplitude $|p|$ is $\sigma_p$, the square root of the polarization variance, and that the probability $P(1/2 < |p|/\sigma_p < 2) = 75\%$, and $P(1/3 < |p|/\sigma_p < 3) = 93.5\%$. Thus the measurement of the polarization in a single field carries cosmological information. From Villumsen (1995), Fig. 1, we see that the implication in an HDM or CDM universe is that the product of $\Omega_0$ and $\sigma_8$ is of order unity. The results in this paper should be confirmed or refuted by other investigations of this and other fields due to the difficult nature of the observations and data reduction.

## Acknowledgments

I am indebted to Jeremy Mould and Todd Small for the acquisition of the data used for this analysis, Ian Smail for data reduction, and Hans Walter Rix, Peter Schneider, and Stella Seitz for helpful suggestions. I am especially indebted to Roger Blandford and Tereasa Brainerd for many vigorous discussions.

## References

Bardeen, J.M., Bond, J.R., Kaiser, N., Szalay, A.S. 1986, ApJ, **304**, 15.
Bertschinger, E., Dekel, A., 1989, ApJ, 336, L5
Blandford, R. D., Saust, A. B., Brainerd, T. G., & Villumsen, J. V., 1991, MNRAS, 251,600 (BSBV)
Dressler, A. *et al.* , 1995, in preparation
Faber, S.M., Burstein, D., 1988, In *Large-Scale Motions in the Universe*, ed. V.C. Rubin and Coyne, G.V., Princeton: Princeton University Press.
Gunn, J. E., 1967, ApJ, 150, 737
Kaiser, N., 1992, ApJ, 388, 272
Kaiser, N., Squires, G., 1993 ApJ, 404, 441
Kristian, J., 1967, ApJ, 147, 864
Miralda-Escudé, J., 1991, ApJ, 380, 1
Mould, J., Blandford, R., Villumsen, J., Brainerd, T., Smail, I., Small, T., & Kells, W., 1994, MNRAS, 271, 31 (M-94)
Valdes, F., 1982, FOCAS Manual, NOAO
Valdes, F., Tyson, J. A., & Jarvis, J. F., 1983, ApJ, 271, 431
Tyson, J.A., Valdes, F., & Wenk, R.A., 1990, ApJ, 349,L1
Villumsen, J.V. 1995, MNRAS, submitted